\documentclass[a4paper,12pt,fleqn]{article}

\usepackage{amsmath}

\newcommand{\pid}{\par\indent}
\newcommand{\pni}{\par\noindent}
\newcommand{\rap}{\left| \dfrac{1-\dfrac{\eta}{2r}}{1+ \dfrac{\eta}
{2r}} \right|}

\begin{document}

\fontsize{12}{18pt}\selectfont

\title{Geometrical Features of  ($4+d$) Gravity}
\author{ A.G.Agnese${}^{\ast}$ and M.La Camera${}^{\dagger}$}
\date{}
\maketitle \vspace{-0.5in}
\begin{center}
{\em Dipartimento di Fisica dell'Universit\`a di Genova\\Istituto
Nazionale di Fisica Nucleare,Sezione di Genova\\Via Dodecaneso 33,
16146 Genova, Italy}
\end{center} \vspace{0.5in}
\begin{abstract}
We obtain the vacuum spherical symmetric solutions for the
gravitational sector of a ($4+d$)-dimensional Kaluza-Klein theory. In
the various regions of parameter space, the solutions can describe
either naked singularities or black-holes or wormholes. We also
derive, by performing a conformal rescaling, the corresponding picture
in the four-dimensional space-time.
\end{abstract} \vspace{0.5in}\pni
PACS numbers: $04.50+h \: ; \: 11.10.Kk$ \vspace{1.5in}\pni
$\overline{\hspace{2in}}$\pni
${}^{*}$E-mail: agnese@ge.infn.it\pni
${}^{\dagger}$E-mail: lacamera@ge.infn.it
\newpage

\section{Introduction}

In recent years, after the great deal of activity since the seventies, the
Kaluza-Klein [1] idea of extra dimensions has
been the subject of revived interest, and the role of compactified
dimensions in physics has appeared in a modern perspective [2].\pid
For instance supersymmetric theories of gravity was particularly
successful in $D=11$: in fact eleven is the maximum number of
dimensions consistent with a single graviton [3] and eleven is also
the minimum number of dimensions required to unify all the forces in
the standard model [4].\pid An appealing aim of Kaluza-Klein theories
in ($4+d$) dimensions, where $d$ is the number of compactified
dimensions, is to construct a pure geometrical description of physics
as a result of minimal extensions to Einstein's theory of general
relativity. It is well known that there are many difficulties in
realizing this goal, nevertheless it seems important to explore the
properties of these higher dimensional theories and to derive
consequences susceptible to experimental test.
For a review of higher dimensional unified theories from the point
of view of general relativity, we refer the reader to the report of
Overduin and Wesson [5]. \pid In this paper we discuss the
gravitational sector of the theory. In sect.$2$ we obtain the
spherically symmetric vacuum solutions in a ($4+d$)-dimensional
space-time. Sect.$3$ is devoted to the classification of these
solutions which, in the various regions of parameter space, can
describe either naked singularities, or black-holes or  wormholes.
In sect.$4$, by a conformal rescaling, we recover the effective
four-dimensional picture and show the equivalence of
($4+d$)-dimensional vacuum general relativity and four-dimensional
general relativity plus a massless scalar field.

\section{The ($4+d$)-dimensional metric tensor}

We consider a ($4+d$)-dimensional vacuum space-time
$M_{4}\bigotimes S_{d}$ , where $M_{4}$ is the ordinary
four-dimensional space-time and $S_{d}$ is a compact internal
manifold with $d$ dimensions.\pid
By introducing coordinates $(x^{\mu},y^{\imath})\quad (\mu = 1,2,3,4 ;
\: \imath = 5,\dots ,4+d)$ where the $y^{\imath}$ have circular
topology $0 \leq y^{\imath} \leq 2\pi\rho^{\imath}$ , and supposing
that the functions depend on the  coordinates $x^{\mu}$only, we can
write the metric in the form
\begin{equation}
ds_{4+d}^{2} = \sum_{\mu ,\nu}g_{\mu ,\nu}(x)dx^{\mu}dx^{\nu} +
\sum_{\imath} C_{\imath}^{2}(x)dy^{\imath}dy^{\imath}
\end{equation}\pid
Because we are not interested here in gauge fields contributions, the
extra off-diagonal terms have been dropped.\pid
The metric tensor will be considered static and three-dimensional
spherically symmetric so, using isotropic coordinates, the
correspondent line element can be written as
\begin{equation}
ds_{4+d}^{2} = -
A^{2}(r)dt^{2}+B^{2}(r)(dr^{2}+r^{2}d\Omega^{2})+\sum_{\imath}
C_{\imath}^{2}(r)dy^{\imath}dy^{\imath}
\end{equation}\pid
The vacuum Einsten equations reduce to the following system of coupled
ordinary differential equations: \bigskip
\begin{flalign*}
&\dfrac{A^{\prime\prime}}{A^{\prime}}+\dfrac{B^{\prime}}{B}+\dfrac{2}
{r}+\sum_{\imath}\dfrac{C_{\imath}^{\prime}}{C_{\imath}}=0 \tag{3a}\\
{}\\ &\dfrac{A^{\prime}}{A}+\dfrac{B^{\prime}}{B}+\dfrac{2}{r}+\dfrac
{C_{\jmath}^{\prime\prime}}{C_{\jmath}^{\prime}}+\sum_{\imath}
\dfrac{C_{\imath}^{\prime}}{C_{\imath}}-\dfrac{C_{\jmath}^{\prime}}
{C_{\jmath}}=0 \qquad (\jmath =5,\dots ,4+d)\tag{3b}\\ {}\\
&\dfrac{B^{\prime\prime}}{B}+\dfrac{B^{\prime}}{B}\left(
\dfrac{A^{\prime}}{A}+\sum_{\imath}\dfrac{C_{\imath}^{\prime}}
{C_{\imath}} \right) +\dfrac{1}{r}\left( \dfrac{A^{\prime}}{A}+3\,
\dfrac{B^{\prime}}{B}+\sum_{\imath}\dfrac{C_{\imath}^{\prime}}
{C_{\prime}} \right)=0 \tag{3c}\\ {}\\
&\dfrac{A^{\prime\prime}}{A}+\dfrac{B^{\prime\prime}}{B}+\sum_{\imath}
\dfrac{C_{\imath}^{\prime\prime}}{C_{\imath}}+\left(
2\,\dfrac{B^{\prime}}{B}+\dfrac{1}{r} \right) \left(
\dfrac{A^{\prime}}{A}+\dfrac{B^{\prime}}{B}+\sum_{\imath}\dfrac
{C_{\imath}^{\prime}}{C_{\imath}} \right)=0 \tag{3d}
\end{flalign*}\pid
\setcounter{equation}{3}
The solutions of system $(3)$ are
\begin{flalign}
&A(r) = \rap^{\dfrac{m}{\eta}} \\
&B(r) = \left( 1+\dfrac{\eta}{2 r} \right) ^{2}\, \rap^{{\displaystyle
1 - \frac{m+\sum_{\imath}\sigma_{\imath}}{\eta}}} \\
&C_{\imath}(r) = \rap^{\dfrac{\sigma_{\imath}}{\eta}}
\end{flalign}
where the parameters $m,\sigma_{\imath}$ and $\eta$ obey the
constraint
\begin{equation}
\eta^{2} = \dfrac{1}{2}\, \left[ m^{2} + \sum_{\imath} \sigma_{\imath}
^{2} + \left( m + \sum_{\imath}\sigma_{\imath} \right) ^{2} \right]
\end{equation}\pid
Thus the line element $(2)$ becomes
\begin{equation} \hspace{-0.65in} \begin{split}
ds_{4+d}^{2} &= -\,\rap^{{\displaystyle
2\,\frac{m}{\eta}}}\hspace{-0.1in}dt^{2}+\left( 1+\dfrac{\eta}{2r}
\right) ^{4} \rap^{{\displaystyle 2\hspace{-0.03in}\left( 1-\frac{m+
\sum_{\imath}\sigma_{\imath}}{\eta} \right)}}\hspace{-0.15in}(dr^{2}+
r^{2}d\Omega^{2})\\ &+\sum_{\imath}\rap^{{\displaystyle 2\,\frac{
\sigma_{\imath}}{\eta}}}dy^{\imath}dy^{\imath}
\end{split}\raisetag{0.5in}\end{equation}\pni
Analizing the behavior of the above line element at spatial
infinity and being aware that in ($4+d$) dimensions the gravitational
coupling constant $G_{4+d}$ is related to the Newton constant $G$ by
\begin{equation}
G = \dfrac{G_{4+d}}{(2\pi)^{d}\prod_{\imath}\rho_{\imath}}
\end{equation}
the parameter $m$ can be identified with the mass and so must be
nonnegative, while each of the quantities $\sigma_{\imath}$ can be of
either sign.\pni
We remind that static and spherically symmetric families of solutions
to five-dimensional relativity have been investigated by other authors
[6],[7],[8].

\section{The ($4+d$)-dimensional picture}

In the isotropic coordinates we used, the standard radial coordinate
$R(r)$ and the radius of compactification $R_{\imath}(r)$ of the
generic $i$-th extra dimension are, respectively, given by
\begin{flalign}
R(r) &= r \left( 1+\dfrac{\eta}{2r} \right) ^{2} \rap^{{\displaystyle
1 - \frac{m+\sum_{\imath}\sigma_{\imath}}{\eta} }}\\
R_{\imath}(r) &= \rho_{\imath}\rap^{\dfrac{\sigma_{\imath}}{\eta}}
\end{flalign}\pid
The requirement that $R(r)$  be a monotonic function
of the radial coordinate $r$ fixes a minimum allowed value
$r_{\text{ min}}$ for $r$ given by
\begin{equation} \hspace{-0.8in} r_{\text{min}} = \begin{cases}
\dfrac{\eta}{2} \hfill \text{for} \quad m+\sum_{\imath}
\sigma_{\imath} < \eta&\\{}\\{\displaystyle \frac{1}{2}
\left[ m+\sum_{\imath}\sigma_{\imath}+ \sqrt{ \left(m+
\sum_{\imath}\sigma_{\imath}\right)^{2} - \eta^{2}}\:\right]} \quad
\text{for} \quad m+\sum_{\imath}\sigma_{\imath} \geq \eta
\end{cases}\end{equation}\pid
One can easily check that the value $r_{\text{min}}$, say $r_{0}$,
which corresponds to the second choice in Eq.(12) never becomes
smaller than the value $\dfrac{\eta}{2}$ and equals it when
$m+\sum_{\imath}\sigma_{\imath} = \eta$.\pid
Let us consider the following cases:\bigskip\pni
$\boldsymbol{a)}$\: $m+\sum_{\imath}\sigma_{\imath} < \eta$\pni
The solutions display, if $m$ is different from zero, infinite
red-shift at $r=\dfrac{\eta}{2}$. Due however to the fact that
$R\left ( \dfrac{\eta}{2} \right ) = 0$, the corresponding surface
area vanishes so we are in front of naked singularities and cannot
even speak of black-holes. On the contrary, each radius $R_{\imath}
(r)$ at $r=\dfrac{\eta}{2}$ either vanishes or equals
$\rho_{\imath}$ or blows up to infinity depending on the value of the
ratio $\dfrac{\sigma_{\imath}}{\eta}$.
This opens the possibility for extra dimensions, which  do not
appear because they are compactified and unobservable at the available
energies, not only to be visible near the naked singularity, but else
to become there the only sensible spatial dimensions.\bigskip\pni
$\boldsymbol{b)}$\: $m+\sum_{\imath}\sigma_{\imath} > \eta$
\bigskip\pni
The line element in Eq.(8) can be written in the standard form as
\begin{equation} \hspace{-0.3in}
ds_{4+d}^{2} = -\,\exp [\,\phi(R)]\, dt^{2}+\displaystyle{
\frac{dR^{2}}{1 - \dfrac{b(R)}{R}}} + R^{2}d\Omega^{2} + \sum_{\imath}
\dfrac{R_{\imath}^{2}(r(R))}{\rho_{\imath}^{2}}\,dy_{\imath}dy_{\imath}
\end{equation}
where
\begin{flalign}
\phi(R) &= \dfrac{m}{\eta}\,\ln{\left| \dfrac{1-\dfrac{\eta}{2r(R)}}
{1+ \dfrac{\eta}{2r(R)}} \right|}\\
1-\dfrac{b(R)}{R} &= \left[\dfrac{r^{2}(R)-(m+\sum_{\imath}
\sigma_{\imath})\,r(R)+\dfrac{\eta^{2}}{4}}{r^{2}(R)-
\dfrac{\eta^{2}}{4}}\right]^{2}
\end{flalign}
and $r(R)$ is the inverse of R(r).\pid
The function $\phi(R)$ is finite everywhere, while the function
$b(R)$ satisfies the conditions $\dfrac{b(R)}{R} \leq 1$ and
$\dfrac{b(R)}{R} \rightarrow 0$ as $R \rightarrow \infty$. Moreover
when $R(r)$ reaches its minimum value $R_{0}=R(r_{0})$ it vanishes the
right-hand side of Eq.(15) and consequently $\dfrac{b(R)}{R}
\rightarrow 1$ as $R \rightarrow R_{0}$.\pid
It then follows that $\phi(R)$ and $b(R)$ play, respectively, the role
of the redshift function and of the shape function of a wormhole, and
$R_{0}$ represents the wormhole throat [9],[10],[11]. The radii of
the extra dimensions remain finite and different from zero.
\bigskip\pni
$\boldsymbol{c)}$\: $m+\sum_{\imath}\sigma_{\imath} = \eta$\pni
If $m$ is different from zero, the solutions display again infinite
red-shift at $r~=~\dfrac{\eta}{2}$, but now $R\left ( \dfrac {\eta}{2}
\right ) = 2\, (m+\sum_{\imath}\sigma_{\imath})$, so the corresponding
surface area does not vanish and represents an event horizon. We
emphasize here the possibility to have black-holes endowed with scalar
charges. If each scalar charge were equal to zero, then the
black-holes would be similar to the Schwarzschild ones.\pni
If $m$ is equal to zero, the solutions represent wormholes
as described above.\pni  Irrespective of $m$, the behavior of each
radius $R_{\imath}(r)$ at $r=\dfrac{\eta}{2}$ depends only on the
value of the ratio $\dfrac{\sigma_{\imath}}{\eta}$.\bigskip\pni

\section{The four-dimensional picture}

 In order to exhibit the physical meaning of the parameters used till
now, we conformally rescale (see for istance [12]) the
($4+d$)-dimensional metric
\begin{equation} \hspace{-0.2in}
ds_{4+d}^{2} = \rap^{-\, \dfrac{\sum_{\imath}\sigma_{\imath}}{\eta}}
\left\{ ds_{4}^{2}+\sum_{\imath} \rap^{\dfrac{2\sigma_{\imath} +
\sum_{\jmath}\sigma_{\jmath}}{\eta}}dy^{\imath}dy^{\imath} \right\}
\end{equation}
where $ds_{4}^{2}$ is the four-dimensional space-time metric:
\begin{equation}\begin{split}
ds_{4}^{2} = &-\, \rap^{\dfrac{2m+\sum_{\imath}\sigma_{\imath}}{\eta}}
dt^{2}\\  &+ \left( 1+\dfrac{\eta}{2r} \right) ^{4}
\rap^{\displaystyle {2-\, \frac{2m+\sum_{\imath}\sigma_{\imath}}
{\eta}}}(dr^{2}+r^{2}d\Omega^{2})
\end{split}\end{equation}\pid
It is apparent that the physical mass $M$ is now given by
\begin{equation}
M = m + \dfrac{1}{2}\,\sum_{\imath}\sigma_{\imath}
\end{equation}
which must be a nonnegative quantity.\pid
Every time an extra dimension $i$ is reduced, a massless scalar
field $\varphi_{\imath}$ is defined by the identity
\begin{equation}
\exp\left\{\,\displaystyle{ 2 \left| \dfrac{2\sigma_{\imath}+
\sum_{\jmath}\sigma_{\jmath}}{\sigma_{\imath}} \right| ^{\frac{1}{2}}
\varphi_{\imath}}\right\} = \rap ^{\dfrac{2\sigma_{\imath}+
\sum_{\jmath}\sigma_{\jmath}}{\eta}}
\end{equation}
whence
\begin{equation}
\varphi_{\imath}(r) = \varepsilon_{\imath} \dfrac{|\sigma_{\imath}
(2\sigma_{\imath}+\sum_{\jmath}\sigma_{\jmath})|^{\frac{1}{2}}}{2\eta}
\,\ln\rap
\end{equation}
Here \, $\varepsilon_{\imath} = \textsl{sign}(2\sigma_{\imath}+
\sum_{\jmath}\sigma_{\jmath})$.\pid
From the four-dimensional Einstein's equations \: $G_{\mu\nu} = 8\pi
T_{\mu\nu}$ \: it is \linebreak straightforward to calculate the
energy-momentum tensor $T_{\mu\nu}$ which can be written as
\begin{equation}
T_{\mu\nu} = \dfrac{1}{4\pi}\,\sum_{\imath} \,\left(\nabla_{\mu}
\varphi_{\imath}\nabla_{\nu}\varphi_{\imath} -\, \frac{1}{2}g_{\mu\nu}
\nabla^{\lambda}\varphi_{\imath}\nabla_{\lambda}\varphi_{\imath}\right)
\end{equation}\pid
If we recall Eq.(20), the energy-momentum tensor acquires the form
\begin{equation}
T_{\mu\nu} = \dfrac{1}{4\pi}\,\left(\nabla_{\mu}\Phi\nabla_{\nu}\Phi -
\dfrac{1}{2}g_{\mu\nu} \nabla^{\lambda}\Phi\nabla_{\lambda}\Phi\right)
\end{equation}
having defined a single effective scalar field $\Phi (r)$ as
\begin{equation}
\Phi (r) = \dfrac{\sigma}{\eta}\, \ln \rap
\end{equation}
and a related scalar charge $\sigma$ given by
\begin{equation}
\sigma^{2} = \sum_{\imath}\, \dfrac{\sigma_{\imath}(2\sigma_{\imath} +
\sum_{\jmath}\sigma_{\jmath})}{4} = \dfrac{2\sum_{\imath}
\sigma_{\imath}^{2}+(\sum_{\imath}\sigma_{\imath})^{2}}{4}
\end{equation}\pid
The parameter $\eta$ defined by Eq.(7) reduces simply to
\begin{equation}
\eta = \sqrt{M^{2}+\sigma^{2}}
\end{equation}
and the line element in Eq.(17) becomes
\begin{equation} \hspace{-0.4in}
ds_{4}^{2} = -\, \rap ^{\displaystyle{2 \frac{M}{\eta}}}\hspace
{-0.1in}dt^{2} + \left( 1+\dfrac{\eta}{2r} \right)^{4} \rap^
{\displaystyle{2\left( 1-\, \frac{M}{\eta}\right)}}\hspace{-0.2in}
(dr^{2}+r^{2}d\Omega^{2})
\end{equation}\pid
This line element, where the singularity at $r = \dfrac{\eta}{2}$ is
naked, was obtained in past years starting from four-dimensional
general relativity plus a massless scalar field (see, for instance,
[13] and references quoted therein).

\section{Conclusions}

In this brief paper we have treated some features of a ($4+d$)
dimensional Kaluza-Klein theory governed by the vacuum field
equations.\pid We found that the higher-dimensional theory can
describe not only black-holes and naked singularities but, what seems
a remarkable propriety, also wormholes. Reducing by a conformal
rescaling to the conventional four-dimensional theory in which gravity
is coupled to a massless scalar field, only naked singularities can
survive because here holds the no-hair theorem and are not violated
the energy conditions which prevent the formation of wormholes
[14].\pid
As a concluding remark, we notice that our solutions would remain
the same if changing the signature of the extra dimensions from
space-like to time-like. It is well known that time-like extra
dimensions give rise to  problems of causality or of insufficient
predictability [15], nevertheless some of these drawbacks might be
overcome by the fact that extra dimensions are compactified [16].


\newpage
\begin{thebibliography}{00}
\bibitem{1}  T.Kaluza, Sitz. Preuss. Akad. Wiss. Phys. Math. K1
966 (1921) \\ O.Klein, Z. Physik \textbf{37}, 895 (1926)
\bibitem{2}  M.J.Duff, NI-94-015, hep-th/9410046
\bibitem{3}  W.Nahm, Nucl. Phys. \textbf{B135}, 149 (1978)
\bibitem{4}  E.Witten, Nucl. Phys. \textbf{B186}, 412 (1981)
\bibitem{5}  J.M.Overduin and P.S.Wesson, Phys. Reports \textbf{283},
303 (1997)
\bibitem{6}  D.J.Gross and M.J.Perry, Nucl. Phys. \textbf{B226}, 29
(1983)
\bibitem{7}  A.Davidson and D.A.Owen, Phys. Lett. \textbf{155B},
247 (1985)
\bibitem{8}  P.S.Wesson, Phys. Lett. \textbf{276B}, 299 (1992)
\bibitem{9}  M.S.Morris and K.S.Thorne, Am. J. Phys. \textbf{56}, 395
(1988)
\bibitem{10}  M.Visser, \emph{Lorentzian Wormholes: From Einstein to
Hawking}, (American Institute of Physics, Woodbury, N.Y., 1995)
\bibitem{11}  A.G.Agnese and M.La Camera, Phys. Rev. D \textbf{51},
2011 (1995)
\bibitem{12}  A.P.Billyard and A.A.Coley, Mod. Phys. Lett. A
\textbf{12}, 21 (1997)
\bibitem{13}  A.G.Agnese and M.La Camera, Phys. Rev. D \textbf{31},
1280 (1985)
\bibitem{14}  D.Hochberg and M.Visser, gr-qc/9802046
\bibitem{15}  M.Tegmark, Class. Quant. Grav. \textbf{15}, 69 (1997)
\bibitem{16}  J.P.Baptista, A.B.Batista and J.C.Fabris, Int. J.
Mod. Phys. D \textbf{2}, 431 (1993)
\end{thebibliography}
\end{document}